\documentclass[a4paper,fleqn,usenatbib]{mnras}

\usepackage{newtxtext,newtxmath}

\usepackage[T1]{fontenc}
\usepackage{ae,aecompl}
\usepackage{hyperref}

\usepackage{graphicx}	
\usepackage{amsmath}	
\usepackage{amssymb}	
\usepackage{adjustbox}
\usepackage{booktabs}	
\usepackage{subfigure}

\newcommand{\vsini}{\ensuremath{{\upsilon}\sin i}}
\newcommand{\kms}{\,km\,s$^{-1}$}

\title{A Spectroscopic and Photometric Investigation of the Mercury-Manganese Star KIC\,6128830}

\author[H{\"u}mmerich et al.]{
Stefan H{\"u}mmerich,$^{1,2}$\thanks{E-mail: ernham@rz-online.de}
Ewa Niemczura,$^{3}$
Przemys{\l}aw Walczak,$^{3}$
Ernst Paunzen,$^{4}$
\newauthor
Klaus Bernhard,$^{1,2}$
Simon J. Murphy,$^{5}$
Dominik Drobek$^{3}$
\\
$^{1}$American Association of Variable Star Observers (AAVSO), Cambridge, USA\\
$^{2}$Bundesdeutsche Arbeitsgemeinschaft f{\"u}r Ver{\"a}nderliche Sterne e.V. (BAV), Berlin, Germany\\
$^{3}$Instytut Astronomiczny, Uniwersytet Wroc{\l}awski, PL-51-622 Wroc{\l}aw, Poland\\
$^{4}$Department of Theoretical Physics and Astrophysics, Masaryk University, Kotl\'a\v{r}sk\'a 2, 611 37 Brno, Czech Republic\\
$^{5}$Sydney Institute for Astronomy (SIfA), School of Physics, The University of Sydney, NSW 2006, Australia\\
}

\date{Accepted XXX. Received YYY; in original form ZZZ}

\pubyear{2017}

\begin{document}

\def\teff{${T}_{\rm eff}$}
\def\kms{{km\,s}$^{-1}$}
\def\logg{$\log g$}
\def\micro{$\xi_{\rm t}$}
\def\macro{$\zeta_{\rm RT}$}
\def\rad{$v_{\rm r}$}
\def\vsini{$v\sin i$}
\def\ebv{$E(B-V)$}
\def\kepler{\textit{Kepler}}
\defcitealias{AGSS09}{AGSS09}

\label{firstpage}
\pagerange{\pageref{firstpage}--\pageref{lastpage}}
\maketitle

\begin{abstract}
The advent of space-based photometry provides the opportunity for the first precise characterizations of variability in Mercury-Manganese (HgMn/CP3) stars, which might advance our understanding of their internal structure. We have carried out a spectroscopic and photometric investigation of the candidate CP3 star KIC\,6128830. A detailed abundance analysis based on newly-acquired high-resolution spectra was performed, which confirms that the star's abundance pattern is fully consistent with its proposed classification. Photometric variability was investigated using four years of archival \kepler\ data. In agreement with results from the literature, we have identified a single significant and independent frequency $f_1$\,=\,0.2065424\,d$^{-1}$ with a peak-to-peak amplitude of $\sim$3.4\,mmag and harmonic frequencies up to $5f_1$. Drawing on the predictions of state-of-the-art pulsation models and information on evolutionary status, we discuss the origin of the observed light changes. Our calculations predict the occurrence of g-mode pulsations at the observed variability frequency. On the other hand, the strictly mono-periodic nature of the variability strongly suggests a rotational origin. While we prefer the rotational explanation, the present data leave some uncertainty.
\end{abstract}




\section{Introduction} \label{introduction}

The chemically peculiar (CP) stars are typically encountered between spectral types early B to early F and make up about 10\,\% of upper main-sequence objects. They exhibit peculiar atmospheric compositions that deviate significantly from the solar abundance pattern. Some CP stars are characterized by large excesses (up to several orders of magnitude) of heavy elements such as Si, Cr, Hg or the rare-earth elements \citep{preston74}, while the atmospheres of other groups of CP stars are depleted in certain elements, as has been established e.g. for the He-weak stars. The chemical peculiarities are generally thought to be caused by atomic diffusion, i.e. the interplay of radiative levitation and gravitational settling operating in the calm radiative atmospheres of slowly rotating stars \citep{michaud70,richer00}. Most elements sink under the influence of gravity; however, those with absorption lines near the local flux maximum are radiatively driven outward.

The CP stars are commonly divided into four groups: CP1 stars (the metallic-line or Am/Fm stars), CP2 stars (the magnetic Bp/Ap stars), CP3 stars (the Mercury-Manganese or HgMn stars) and CP4 stars (the He-weak stars). The CP2/4 stars are characterized by the presence of globally-organized magnetic fields, which are likely of fossil origin \citep{braithwaite04} and attain strengths of up to several tens of kG \citep{auriere07}. The CP1/3 stars are generally considered to be non-magnetic or only very weakly magnetic objects. However, several recent studies have challenged this view, as discussed below.

The CP3 stars, which are relevant to the present investigation, occupy the temperature range of 10\,000\,K\,$\leq$\,\teff\,$\leq\,$16\,000\,K \citep{smith96, ghazaryan16} and are generally very slow rotators. They have no or only very little convective motion in their atmospheres and exhibit unusually strong surface enhancements of Hg and Mn (up to 6 and 3 dex, respectively). Numerous other peculiarities are observed in these stars, such as increased abundances of elements like P, Y, Sr and Xe, and depletions of other elements like He, Ni or Al \citep{castelli04}. Heavy elements are generally found to be overabundant; the strength of the observed overabundance typically increases with atomic number. CP3 stars present highly individualistic abundance patterns, and striking differences in abundance ratios are observed from one star to another \citep{ghazaryan16}. It has yet to be fully understood why diffusion processes are creating significantly different atmospheric compositions in stars of similar temperature \citep{cowley14}. A binarity fraction of more than 50\,\% (up to 91\,\%; \citealt{schoeller10}) has been established among the CP3 stars \citep{smith96} and has been proposed to have a fundamental impact on the observed chemical peculiarities, primarily through tidal interaction.

At least some CP3 stars exhibit an inhomogeneous distribution of elements on their surfaces, and obvious signs of secular evolution of these structures have been found \citep{kochukhov07}. Variations of the line profiles of several atomic species have been observed and interpreted in terms of chemical spots of elements such as Ti, Sr and Y, which are assumed to be present at high atmospheric altitudes \citep{adelman02,hubrig06,briquet10}. This is reminiscent of the magnetic CP2 stars, which are characterized by the presence of surface abundance inhomogeneities. Flux is redistributed in the chemical spots, and -- as the stars rotate -- the changing viewing angle leads to photometric variability with the rotation period \citep{stibbs50}. While spot formation in CP2 stars is quite well understood and related to the presence of strong, globally-organized magnetic fields, the situation is less clear for the CP3 stars, which have traditionally been regarded as lacking strong, organized magnetic fields.

Considerable effort has therefore been devoted to the investigation of the magnetic properties of CP3 stars, with conflicting results. Several studies have indicated the presence of weak or tangled fields in these objects \citep{hubrig10,hubrig12}, while other studies failed to detect magnetic field signatures in the spectral line profiles (see \citealt{kochukhov13}, and references therein). To explain the observed abundance inhomogeneities, several mechanisms (only some of which depend on the presence of magnetic fields) have been proposed and investigated, e.g. time-dependent atomic diffusion \citep{alecian09} and atomic diffusion in the presence of a weak, multipolar magnetic field \citep{alecian12,alecian13}.

CP3 stars occupy a part of the Hertzsprung-Russell (HR) diagram that also harbours pulsating variables such as the (\lq chemically-normal\rq) Slowly-Pulsating B (SPB) stars. Indeed, \citet{turcotte03} have presented calculations predicting an iron accumulation in the upper layers of the envelopes of CP3 stars, which should lead to pulsational driving via the $\kappa$ mechanism and variability similar to that observed in SPB stars. However, observations have largely contradicted theoretical predictions: (single) CP3 stars are among the least photometrically variable of the CP stars \citep{adelman98}, which suggests that important physical processes are not covered adequately by the models.

More recently, \citet{alecian09} have also presented calculations that favour the excitation of non-radial g modes in CP3 stars, as found in SPB stars. Drawing on precise photometry from the \textit{CoRoT} space mission \citep{Corot}, they were able to establish periodic light variations with properties compatible with the theoretically-predicted pulsational variability (periods of 2.53 and 4.3\,d, amplitudes less than 1.6\,mmag) in at least two of the three investigated faint CP3 stars. The authors presented arguments that favour pulsational variability as the source of the observed photometric variability but were not able to reach a final conclusion.

In their investigation of the variability of B-type stars using early \kepler\ data \citep{Kepler} from quarters Q0 to Q4, \citet{balona11} identified nearly sinusoidal, monoperiodic light variations with a period of 4.84\,d and an amplitude of 3.6\,mmag in the CP3-star candidate KIC\,6128830 \citep{catanzaro10}. They did not investigate the star in detail; however, based on the properties of the photometric variability and the consistency of the photometric period with the \vsini\ value of 15\,\kms\ determined by \citet{catanzaro10}, they suggested rotational modulation as the underlying mechanism of the observed light variations.

Using observations from the \textit{STEREO} spacecrafts \citep{Stereo}, \citet{paunzen13} identified light variations in 11 CP3 stars. For seven objects, photometric variability was established for the first time. However, because of a lack of data, no decision about the type of variability could be reached, and the authors cautioned that further observations are needed. Detailed analyses for these objects are not yet available.

Using \textit{CoRoT} observations, \citet{ghazaryan13} investigated the bright CP3 star HD\,175640, with the main aim of probing the existence of granulation signature. In agreement with the expectations, no granulation signature was found. Furthermore, no clear pulsational signal was identified. The authors conclude that, even if pulsational variability may be present in some CP3 stars, it is not a systematic characteristic of the group of CP3 stars as a whole.

\citet{morel14} have reviewed the current status of photometric variability in CP3 stars and presented a detailed investigation of the light variability of the CP3 star HD\,45975 using \textit{CoRoT} observations, with the aim of investigating, and possibly establishing, the presence of pulsations in this object. They have found significant differences in the level of variability observed in the spectral lines from one element to the next. While obvious periodic variations are present in the line profiles of elements such as Y, Mn or Cr, no variability has been detected for other elements such as Hg. Furthermore, a phase lag between the variability of the different ions was established. The authors concluded that the light variability of HD\,45975 is more compatible with rotational modulation of abundance spots on the surface and, if this assumption is correct, place an upper limit of 50\,ppm on the amplitude of any possible pulsational variations in HD 45975.

Further evidence for the existence of rotational variability in CP3 stars has been coming forth recently. \citet{strassmeier17} identified complex out-of-eclipse variability in the eclipsing, double-lined spectroscopic binary system HSS\,348. At least the primary component was found to be a CP3 star. The out-of-eclipse variability is characterized by four nearly equidistant minima of different depth that remained stable in shape and amplitude throughout the \textit{CoRoT} observations. Their individual photometric periods are harmonics of the same fundamental period ($P$\,=\,12.47\,d) which agrees, within the errors, with the transit period. The authors concluded that the out-of-eclipse variability is caused by rotational modulation due to an inhomogeneous surface structure.

\kepler\ K2 photometry of the Pleiades has revealed mono-periodic variability at a 10.29-d period in Maia (HD\,23408; spectral type B7\,III; \citealt{white17}). Phase-resolved spectroscopy showed the variability is compatible with the oblique rotator model. Since Maia has been alternately classified as a CP3 and CP4 star (see the discussion in \citealt{white17}), with weak He, strong Mn, but normal Hg lines, it may represent an intermediate object between both classes. Hence Maia is an important piece in this puzzle, but it is not yet clear where it fits.

In summary, there is no concensus yet as to what causes the observed photometric variations in CP3 stars. Arguments have been presented in favour of the existence of rotational as well as pulsational variability in these objects, with the more recent studies favouring rotational modulation. In order to add another piece to this puzzle, which might eventually contribute to its solution, we here present a detailed spectroscopic and photometric analysis of the CP3-star candidate KIC\,6128830 (BD+41\,3394, HIP 96210, 2MASS J19334968+4128452; RA,\,Dec\,(J2000) = 19h\,33m\,49s.685, +41$^\circ$\,28'\,45''.24 \citep{vanleeuwen07}; $V$\,=\,9.2\,mag \citep{hog00}), which has not yet been the subject of a detailed investigation, with the aim of confirming its suspected status as a CP3 star and unraveling the mechanism behind its photometric variability. To this end, new spectroscopic observations and all four years of \kepler\ photometry have been employed.


\section{Spectroscopic analysis} \label{spectroscopy}

In their spectroscopic characterization of \kepler\ early-type targets, \citet{catanzaro10} identified KIC\,6128830 as a CP3 star candidate based on the derived values of \teff\ and \logg\ and its chemical peculiarities. In addition to a strong ($\sim$2.2\,dex) overabundance of Mn, they identified a general overabundance of C, O and all iron-peak elements. However, the authors cautioned that this classification needs to be confirmed by additional spectroscopic material because their observations did not cover the blue spectral range containing the Hg line at $3984$\,{\AA} -- an important diagnostic for CP3 stars. We have obtained new spectroscopic material in order to investigate KIC\,6128830 in more detail and confirm its suspected status as a CP3 star.

\begin{figure*}
  \centering
    \includegraphics[width=0.9\textwidth]{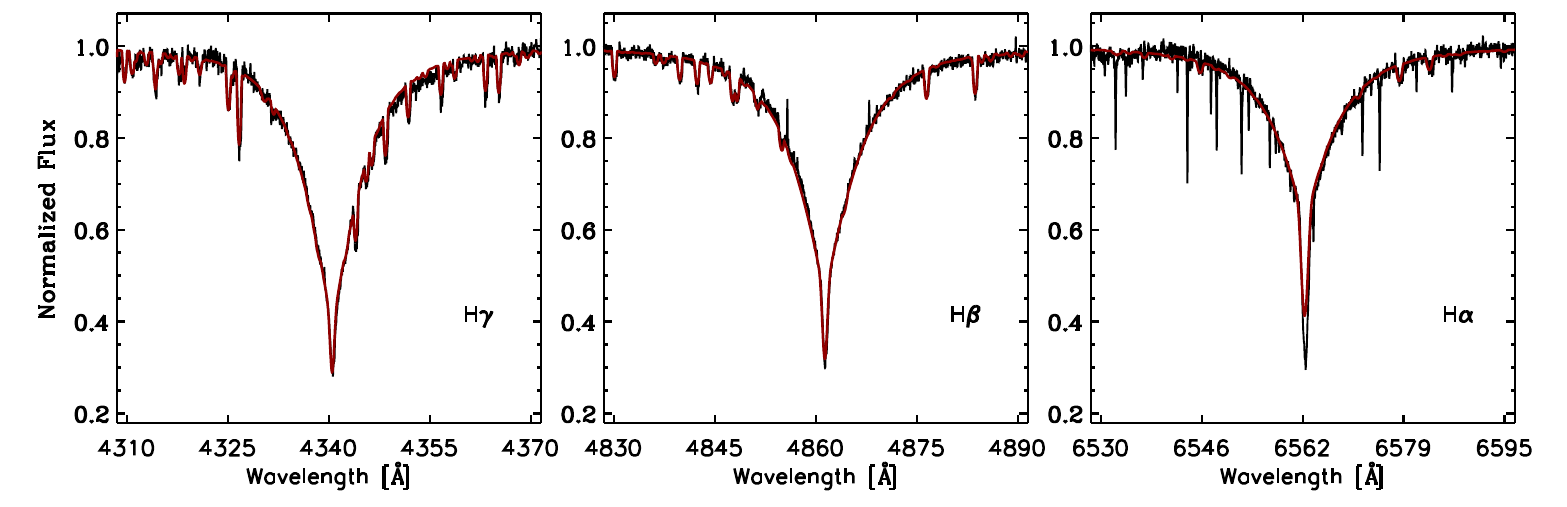}
    \caption{Observed hydrogen H$\gamma$, H$\beta$ and H$\alpha$ lines (black) and best-fit synthetic profiles (red).}
    \label{balmer}
\end{figure*}

\subsection{Spectroscopic observation}

The investigated spectrum of KIC\,6128830 was taken with the cross-dispersed, fibre-fed \'{e}chelle spectrograph HERMES \citep[High Efficiency and Resolution Mercator \'{E}chelle Spectrograph,][]{raskin} attached to the $1.2$-m Mercator telescope located on La Palma (Canary Islands). The spectrum with resolving power ${\rm R}\sim85\,000$ covers the visual part from $3770$ to $9000$\,{\AA}, with signal-to-noise (S/N) ratio equal to $90$ at $5500$\,{\AA} and was taken at BJD\,=\,2455757.729 (midpoint of observation). It has been reduced with a dedicated pipeline\footnote{\url{http://www.mercator.iac.es/instruments/hermes/drs/}} \citep{raskin11}, which includes all necessary steps, namely bias subtraction, extraction of scattered light produced by the optical system, cosmic ray filtering, division by a normalized flat-field, wavelength calibration by a ThArNe lamp and order merging. The normalisation to the continuum was performed with the standard {\sc iraf}\footnote{Image Reduction and Analysis Facility, \url{http://iraf.noao.edu/}} procedure {\it continuum}.

\subsection{Spectral classification}

We performed a spectral classification of our star following current best practice and updated spectral standards \citep{gray09}. Usually, the spectral classification process of normal B-type stars relies mainly on H and He lines. Since we have noticed that the spectrum of KIC\,6128830 shows He lines slightly weaker than expected for normal B stars, the He lines were excluded from the classification process. Fortunately, for late B-type stars, hydrogen lines are sensitive to both luminosity and temperature. Therefore, the spectral classification of KIC\,6128830 was carried out on the basis of the Balmer lines H$\gamma$ and H$\beta$ and yielded a spectroscopic class of B7\,III. Due to the poor quality of the spectrum in this spectral range, the H$\delta$ line was not suitable for analysis.

In the next step, the spectrum was scanned for the presence of peculiarities. In addition to the He weakness, a strong Hg line (Hg\,II at $\lambda$\,3984) and several enhanced Mn lines (e.g. $\lambda$\,4136, $\lambda$\,4206 and $\lambda$ 4252) were easily detected, which points to the star being a typical CP3 object. Our final classification of KIC\,6128830 as B7\,III HgMn is consistent with the results of \citet{catanzaro10}.

\subsection{Atmospheric data}

The \kepler\ Input Catalogue (KIC; \citealt{brown11}) lists an effective temperature of \teff\,$=9778$\,K and surface gravity of \logg\,$=4.315$ for our programme star, as determined from photometric calibrations. In the \citet{huber14} catalogue of revised stellar properties of \kepler\ targets, two sets of photometrically derived parameters are given: \teff\,$=10001\pm350$\,K and \logg\,$=4.315\pm0.400$, and \teff\,$=9990^{+303}_{-381}$\,K and \logg\,$=4.201^{+0.115}_{-0.394}$.

A medium-resolution (${\rm R} = 21 000$) FRESCO spectrum of KIC\,6128830 was analyzed by \citet{catanzaro10}. The relatively narrow spectral range ($\lambda\lambda$ $4300-6800$) did not allow for full spectral classification, but \teff\ and \logg\ were determined by minimizing the difference between the observed and the synthetic H$\beta$ profiles. As results, \teff\,$=12600\pm600$\,K and \logg\,$=3.5\pm0.3$ were obtained.

The same method of utilizing the sensitivity of the hydrogen lines to effective temperature and surface gravity was used here to determine these atmospheric parameters. From an investigation of the H$\gamma$, H$\beta$ and H$\alpha$ lines, we have derived \teff\,$=13000\pm400$\,K and \logg\,$=3.5\pm0.15$, in agreement with the results of \citet{catanzaro10}. Again, due to the poor quality of the spectrum in this range, the H$\delta$ line was not considered. Similar to the aforementioned investigators, we used an iterative approach which minimizes the differences between the observed and synthetic Balmer line profiles. To estimate the error of \teff\ and \logg, we took into account the differences in the obtained \teff\ and \logg\ values from separate Balmer lines, resulting from validity of normalization. Furthermore, from an analysis of Fe\,I and Fe\,II lines, we obtained a microturbulence of \micro\,$=0.6\pm0.2$\,\kms\ and a projected rotational velocity of \vsini\,$=25\pm2$\,\kms. The \vsini\ value determined in this investigation is higher by $10$\,\kms\ than the value derived by \citet{catanzaro10} from mid-resolution spectrum analysis. Fig.\,\ref{balmer} illustrates the observed H$\gamma$, H$\beta$ and H$\alpha$ lines and the synthetic profiles calculated for the final atmospheric parameters. The synthetic H$\gamma$ and H$\beta$ lines fit the observations well. The discrepancy between the synthetic and observed core of H$\alpha$ is the result of assuming local thermodynamic equilibrium (LTE) in our analysis. For H$\alpha$, departures from LTE produce a deeper line-core and shallower wings \citep{mihalas72}. Nevertheless, the best fit to all three lines (respectively the wings in the case of H$\alpha$) yield similar values.

The radial velocity was determined comparing the observed spectrum to the synthetic one for the final atmospheric parameters and abundances. To this end, we compared various metal lines in the whole frame, after which an average value was calculated. From our spectrum, we derive $RV$\,$=-4.2\pm1.6$\,\kms\, which differs from the value derived by \citet{catanzaro10} ($RV$\,$=+6.6\pm1.0$\,\kms). This may possibly point to the presence of a companion star. However, further RV measurements are necessary to tackle this question, which is beyond the scope of the present investigation.

The employed atmospheric models (1-dimensional, plane-parallel, hydrostatic and radiative equilibrium) were calculated with the {\sc ATLAS\,9} code \citep{kurucz14}. The grid of models was calculated for effective temperatures from $10000$ to $15000$\,K, with a step size of $100$\,K; surface gravities from $3.0$ to $4.3$\,dex, with a step size of $0.1$\,dex; microturbulence velocities between $0.0$ and $1.0$\,km\,s$^{-1}$; and for metallicities [M/H] equal to $0.0$, $\pm0.5$ and $\pm1.0$\,dex. The synthetic spectra were computed with the Fortran {\sc SYNTHE} code \citep{kurucz05}, which calculates intensity stellar spectra for a given model atmosphere under the assumption of LTE. Both codes have been ported to GNU/Linux by \citet{sbordone05}. We used the line lists from Fiorella Castelli's website\footnote{http://wwwuser.oats.inaf.it/castelli/linelists.html} that were introduced in \citet{castelli04}. The lists were subsequently updated with data from Robert Kurucz's website\footnote{http://kurucz.harvard.edu/linelists/gfnew/ (October 2016 version)}. The atomic data were supplemented for the second and third spectra of the lanthanides with the data taken from the Vienna Atomic Line Database \citep[VALD,][]{kupka99}, originally presented in the Data on Rare Earths At Mons University (DREAM) database \citep{biemont99}.

\subsection{Chemical abundances} \label{abundances}

\begin{figure*}
  \centering
    \includegraphics[width=0.9\textwidth]{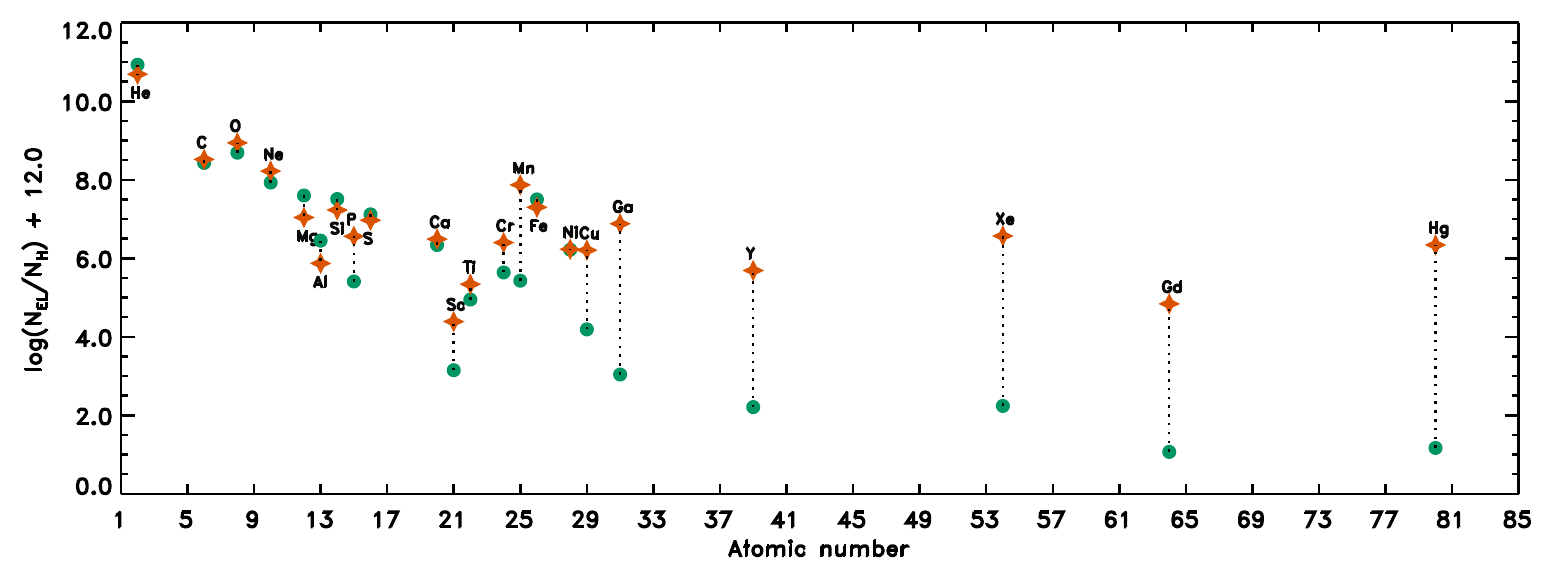}
    \caption{Comparison of the chemical composition of KIC\,6128830 (orange stars) to the solar abundance pattern (green circles). Overabundant and underabundant elements in KIC\,6128830 are indicated, respectively, above and below their corresponding symbols.}
    \label{abundances_solar}
\end{figure*}

\begin{figure*}
  \centering
    \includegraphics[width=0.9\textwidth]{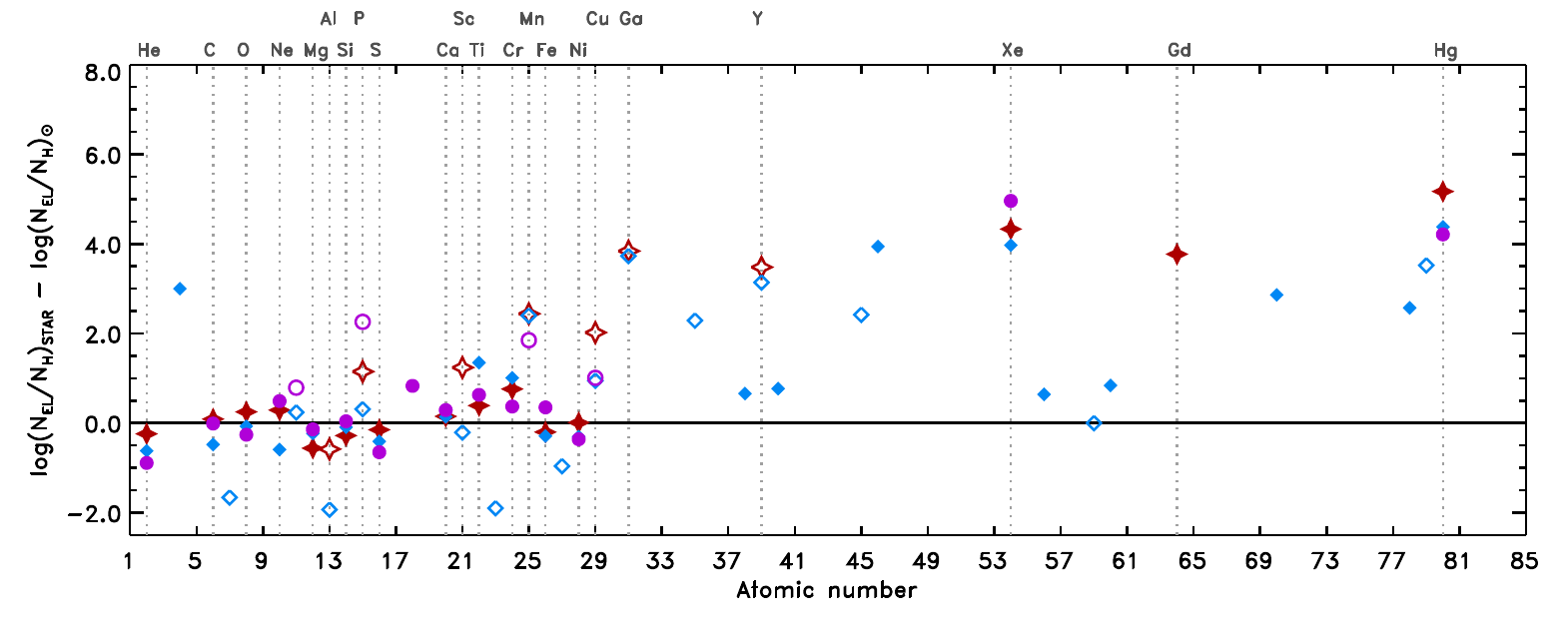}
    \caption{Comparison of the chemical composition of KIC\,6128830 (red stars) to the abundance patterns observed in the CP3 stars HD\,49606 (pink circles) and HD\,175640 (blue diamonds). Odd-$Z$ elements and even-$Z$ elements are denoted by, respectively, open and filled symbols.}
    \label{abundances_hgmn}
\end{figure*}

To derive the chemical abundance pattern of KIC\,6128830, we analyzed the metal lines using spectrum synthesis, which relies on an efficient least-squares optimization algorithm \citep[e.g.][]{niemczura15}. In Table\,\ref{chemical_abundances} and Fig.\,\ref{abundances_solar}, the obtained chemical abundances of KIC\,6128830 have been compared with the photospheric solar values given by \citet[][hereafter AGSS09]{AGSS09}.

Our abundance analysis fully confirms the chemical peculiarities indicated by the spectral classification. Fig.\,\ref{abundances_hgmn} compares the abundance pattern of KIC\,6128830 to the abundance patterns of the two well-known CP3 stars HD\,49606 \citep{catanzaro16} and HD\,175640 \citep{castelli04}. The atmospheric parameters of both these stars are similar to the values derived here for KIC\,6128830. For HD\,175640, \citet{castelli04} derived \teff\,=\,$12000$\,K and \logg\,=\,$3.95$\,dex from an analysis of Str{\"o}mgren $c_1$ and $\beta$ indices and assumed a microturbulence of \micro\,=\,$0.0$\,\kms. For HD\,49606, \citet{catanzaro16} obtained \teff\,$=13000\pm150$\,K and \logg\,$=3.80\pm0.05$ from Balmer line analysis; the value of \micro\ was obtained from Fe lines as $0.3^{+0.9}_{-0.0}$\,\kms. HD\,175640 is rotating very slowly at \vsini\,$=2.5$\,\kms \citep{castelli04}, whereas the projected rotational velocity of HD\,49606 equals $19.0\pm0.5$\,\kms\ \citep{catanzaro16}.

A detailed discussion of the abundance anomalies observed in CP3 stars has been given by \citet{ghazaryan16}, who present an exhaustive review of the recent literature. In the following, the peculiarities of KIC\,6128830 are discussed in the light of this information.

The He lines in the spectrum of KIC\,6128830 are weaker than for chemically-normal stars of the same effective temperature. Most He lines ($\lambda\lambda$ 4120, 4713, 5015, 5875, 6678) can be well modeled with LTE analysis and indicate a He abundance of $10.69 \pm 0.05$\,dex ([He/H]\,=\,$-0.24$\,dex). The remaining He lines ($\lambda\lambda$ 4026, 4471 and 4921) are diffuse-series lines with overlapping forbidden components and have broad and blended wings that cannot be well reproduced with LTE analysis. (For a thorough discussion of the line-broadening theory of He lines, we refer the reader to \citet{auer73}). \citet{przybilla08} analysed the chemical composition of the hyper-velocity star HVS\,7 with similar basic parameters to our target star (\teff\,$=12000\pm500$\,K and \logg\,$=3.8\pm0.1$). They compared LTE and non-LTE abundances of some elements, including He. Non-LTE and LTE abundances obtained from the weak He\,I lines show differences up to $0.3$\,dex, wherein the lower He abundance was derived from non-LTE analysis.

C, O and Ne abundances are close to the solar values \citep{ghazaryan16}, which is typical for CP3 stars. Likewise, the observed underabundances of Mg, Al, Si and S, and the near-solar abundance of Ca, agree well with the values obtained for similar stars \citep{ghazaryan16}. KIC\,6128830 shows strongly enhanced P lines, which is another characteristic feature of CP3 stars, as is the observed odd-Z anomalous abundance pattern in the Si-P-S triplet. The obtained overabundance of Ne could be due to non-LTE effects \citep{dworetsky00}. 

In CP3 stars, the abundances of iron-peak elements are broadly scattered around the solar values. We found Fe to be slightly underabundant in KIC\,6128830, and an odd-Z anomalous abundance pattern is observed in the Cr-Mn-Fe triplet, which agrees with the expectations for CP3 stars \citep{ghazaryan16}. Sc is overabundant in KIC\,6128830, but the value was obtained from the analysis of one weak blend only and should therefore be regarded with caution. Ti, Cr and Mn are overabundant, too; the Mn abundance is more than $2$\,dex higher than the solar value. An enhanced abundance of Cu is observed, in accordance with the expectations.

Ga shows a significant enhancement by $\sim3.8$\,dex. Typically, the Ga overabundance for HgMn stars reaches $4$\,dex. All the considered elements with $Z>30$, namely Y, Xe, Gd and Hg are overabundant, which agrees with the results for other peculiar stars of this type. Finally, the Hg overabundance exceeds $5$\,dex.

In summary, our abundance analysis confirms that the abundance pattern of KIC\,6128830 is fully consistent with the pattern of CP3 stars.

\begin{table}
\begin{tabular}{|l|l|l|l|l|}
\hline
\hline
\toprule
Element  &   $Z$  & Abundance$^{1}$          & No. of     & Solar     \\
         &        &                          & lines used & abundance$^{1}$ \\
\hline
\midrule
{\bf He} &   5    & $10.69 \pm 0.05$  &  5 & $ 10.93 \pm 0.01 $\\
{\bf C } &   6    & $ 8.52 $          &  1 & $  8.43 \pm 0.05 $\\
{\bf O } &   8    & $ 8.94 \pm 0.20$  &  3 & $  8.69 \pm 0.05 $\\
{\bf Ne} &  10    & $ 8.22 \pm 0.11$  &  3 & $  7.93 \pm 0.10 $\\
{\bf Mg} &  12    & $ 7.04 \pm 0.18$  &  5 & $  7.60 \pm 0.04 $\\
{\bf Al} &  13    & $ 5.87 $          &  1 & $  6.45 \pm 0.03 $\\
{\bf Si} &  14    & $ 7.23 \pm 0.11$  & 10 & $  7.51 \pm 0.03 $\\
{\bf P } &  15    & $ 6.56 \pm 0.20$  & 13 & $  5.41 \pm 0.03 $\\
{\bf S } &  16    & $ 6.97 \pm 0.28$  & 23 & $  7.12 \pm 0.03 $\\
{\bf Ca} &  20    & $ 6.49 \pm 0.20$  &  2 & $  6.34 \pm 0.04 $\\
{\bf Sc} &  21    & $ 4.39 $          &  1 & $  3.15 \pm 0.04 $\\
{\bf Ti} &  22    & $ 5.34 \pm 0.19$  & 18 & $  4.95 \pm 0.05 $\\
{\bf Cr} &  24    & $ 6.40 \pm 0.13$  & 16 & $  5.64 \pm 0.04 $\\
{\bf Mn} &  25    & $ 7.87 \pm 0.16$  & 91 & $  5.43 \pm 0.04 $\\
{\bf Fe} &  26    & $ 7.30 \pm 0.19$  & 66 & $  7.50 \pm 0.04 $\\
{\bf Ni} &  28    & $ 6.23 \pm 0.10$  &  5 & $  6.22 \pm 0.04 $\\
{\bf Cu} &  29    & $ 6.21 $          &  1 & $  4.19 \pm 0.04 $\\
{\bf Ga} &  31    & $ 6.88 \pm 0.10$  &  2 & $  3.04 \pm 0.09 $\\
{\bf Y } &  39    & $ 5.69 \pm 0.25$  & 16 & $  2.21 \pm 0.05 $\\
{\bf Xe} &  54    & $ 6.57 $          &  1 & $  2.24 \pm 0.06 $\\
{\bf Gd} &  64    & $ 4.84 \pm 0.25$  & 10 & $  1.07 \pm 0.04 $\\
{\bf Hg} &  80    & $ 6.34 \pm 0.10$  &  4 & $  1.17 \pm 0.08 $\\
\bottomrule
\hline
\multicolumn{5}{l}{$^{1}$ given in $\log(N\textsubscript{EL}/N\textsubscript{H})+$12.0}
\end{tabular}
 \caption{Chemical abundances and standard deviations for individual elements. Solar abundances were taken from \citetalias{AGSS09}.}
 \label{chemical_abundances}
\end{table}


\section{Light curve analysis}

The following sections give an overview over the data source and the employed methods of period analysis.


\subsection{Kepler observations}

The main aim of the \kepler\ mission is the detection of planets around other stars via the transit method, in order to determine the frequency of Earth-like planets in or near the habitable zone of Sun-like stars \citep{Kepler}. The long, quasi-uninterrupted and ultra-precise \kepler\ time series are also used for asteroseismology, which allows us to probe the interior of stars and determine fundamental stellar properties that are otherwise extremely difficult to measure, such as stellar age \citep{aerts15}. For more information on the \kepler\ spacecraft, we refer to \citet{Kepler} and \citet{koch10}.

\kepler\ Simple Aperture Photometry (SAP) data are known to suffer from strong instrumental trends (caused by e.g. aperture changes, long-term drifts and other effects), which are largely compensated for in the Pre-search Data Conditioned (PDC) flux \citep{LCdata,murphy14}. Several algorithms are employed in this respect, such as \lq msMAP\rq, a wavelet-based, band-splitting framework for removing systematics \citep{stumpe14}.

KIC\,6128830 was observed by the \kepler\ satellite during quarters Q0 to Q17, which corresponds to a time span of $\sim$4 years (May 2, 2009 to May 11, 2013; 1470.462 days; duty cycle of 90.75\,\%). A total of 112,258 datapoints were collected, most of which (69,778 data points) were acquired in the long-cadence (LC; 29.4 min-sampling) mode \citep{LCdata}. The remaining 42,480 data points were acquired during Q4 in the short-cadence (SC; $\sim$1 min-sampling) mode \citep{SCdata}. We downloaded the PDC data for KIC\,6128830 from the Mikulski Archive for Space Telescopes (MAST).\footnote{https://archive.stsci.edu/kepler/}


\subsection{Data processing and frequency analysis} \label{section_periodanalysis}

\begin{figure}
\begin{center}
\includegraphics[width=0.47\textwidth]{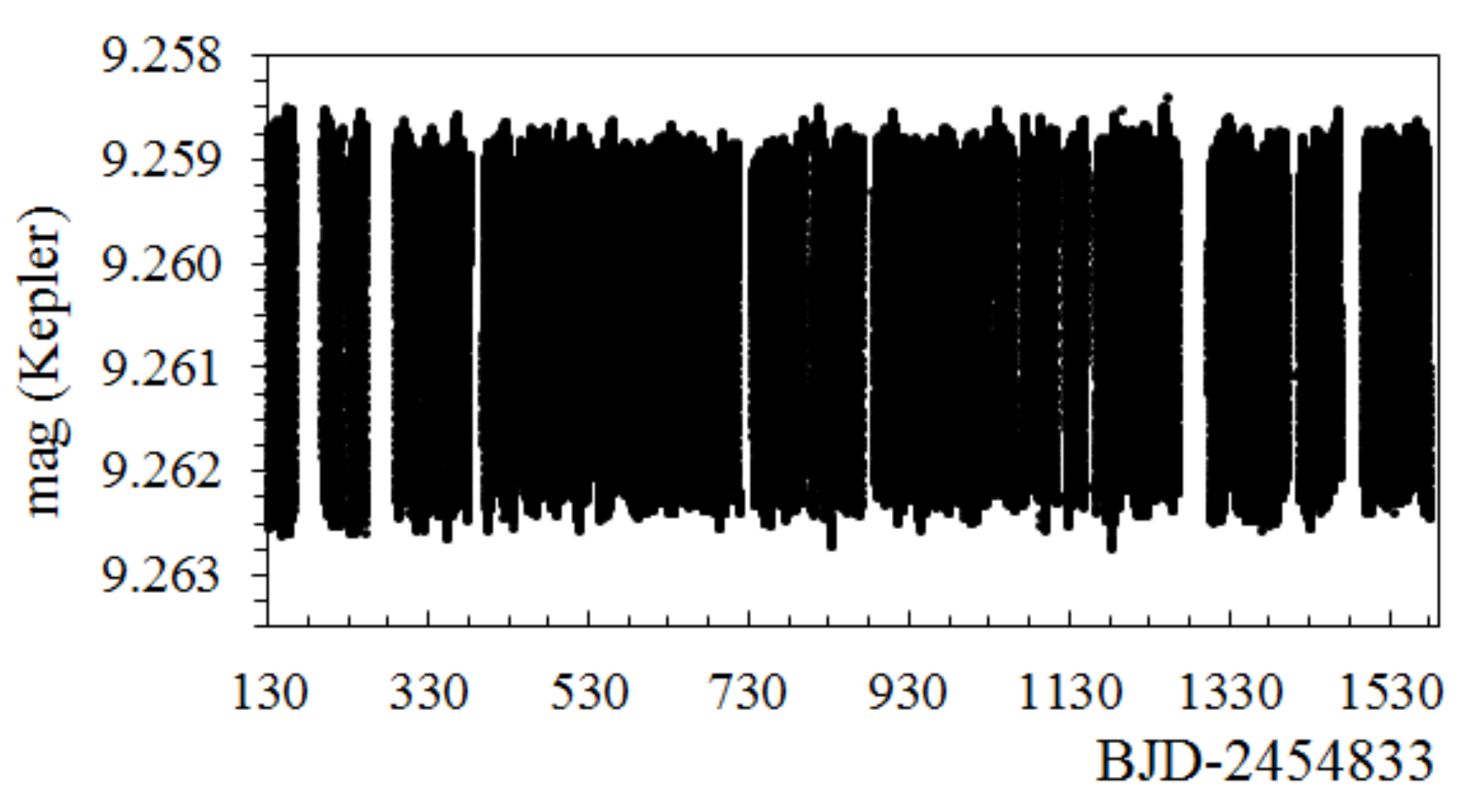}
\includegraphics[width=0.47\textwidth]{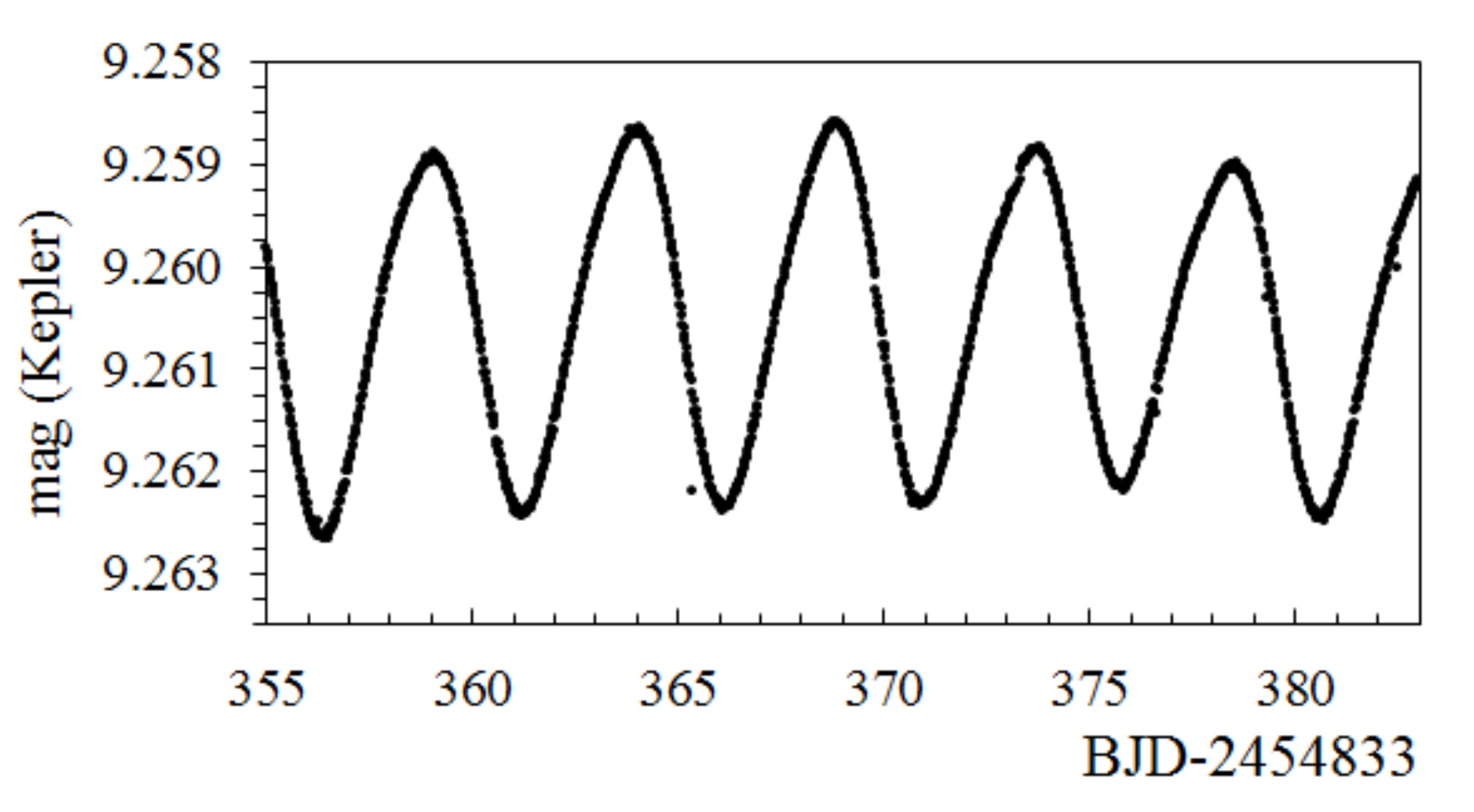}
\caption{Light curve of KIC\,6128830, based on \kepler\ PDC flux procured from the MAST archive. Obvious outliers were removed by visual inspection, and low-order polynomials were fitted and subtracted to reduce the inherent long-term trends. The upper panel illustrates the full light curve, the lower panel shows a detailed view of the light curve that has been based on part of Q3 data ($355<(BJD-2454833)<383$).}
\label{lightcurve}
\end{center}
\end{figure}

The data for KIC\,6128830 suffer from instrumental trends and artefacts. For instance, changes in mean magnitude level are observed, in particular between the corresponding quarters. While these trends appear much reduced in the PDC flux, they are still significant and result in considerable noise at low frequencies. In order to augment this situation, low-order polynomials were fitted and subtracted to reduce the long-term trends. Furthermore, a few obviously outlying data points ($N$\,=\,11) were removed by visual inspection. The resulting light curve, which was used as the basis for our frequency analysis, is shown in Figure \ref{lightcurve}.

In order to test the validity of our approach, the frequency analysis outlined below was repeated using the original dataset as input. While the low-frequency noise appears much reduced in the pre-processed dataset, no significant differences were found in the derived variability frequencies, which makes us confident of the applicability of our detrending procedure.

For the frequency analysis, the software package \textsc{Period04} was employed, which uses discrete Fourier transformation and allows least-squares fitting of multiple frequencies to the data \citep{period04}. In order to extract all relevant frequencies, the data were searched for periodic signals and consecutively prewhitened with the most significant frequency. Periodogram features and phased light curves were visually inspected to prevent instrumental signals from contaminating our results.

An initial search in the range of 0\,<\,$f($d$^{-1})$\,<\,50 was carried out to check for the presence of high-frequency signals. None were found, and the search range was subsequently restricted to 0\,<\,$f($d$^{-1})$\,<\,2. After the identification of the two most significant peaks, the search range was restricted to 0.1\,<\,$f($d$^{-1})$\,<\,2, in order to exclude the low-frequency noise and detect signals of lower amplitudes.

\begin{table}
\begin{center}
\caption{Significant frequencies, semi-amplitudes and signal-to-noise ratios detected for KIC\,6128830, as derived with \textsc{Period04}.}
\label{pa_table}
\begin{tabular}{cllr}
\hline
\hline
ID & frequency  & semi-amp. & S/N\\
   & [d$^{-1}$] & [mmag] & \\       
\hline
$f_1$	& 0.2065424 & 1.701 & 186.0 \\
$2f_1$ & 0.413083 & 0.187 & 63.8 \\
$3f_1$ & 0.619720 & 0.019 & 11.2 \\
$4f_1$ & 0.826099 & 0.008 & 6.4 \\
$5f_1$ & 1.032896 & 0.006 & 5.6 \\
\hline
\end{tabular}
\end{center}
\end{table}

\begin{figure}
\begin{center}
\includegraphics[width=0.47\textwidth]{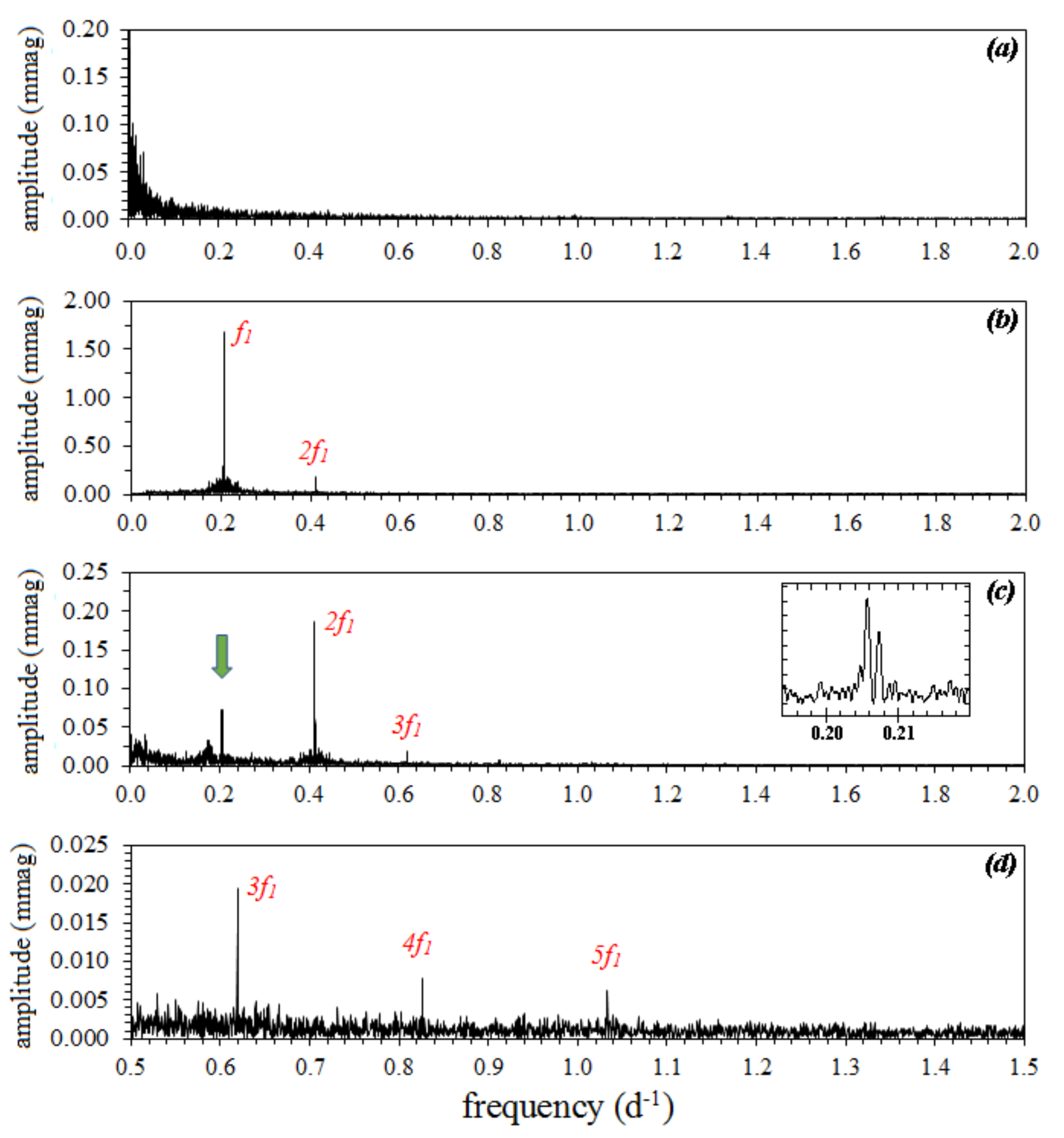}
\caption{Period analysis based on Kepler data for KIC\,6128830, illustrating important steps of the frequency spectrum analysis. The top panel (a) shows the spectral window dominated by low-frequency noise. The other panels illustrate the frequency spectra for (b) unwhitened data, (c) data that has been prewhitened with $f_1$ and (d) data that has been prewhitened with $f_1$ and $2f_1$. Note the different scales on the ordinates and the abscissa in panel (d). The green arrow in panel (c) indicates the position of the residual peaks of the main frequency $f_1$. A detailed view of the corresponding part of the frequency spectrum is provided in the inset. The ordinate axes denote semi-amplitudes, as derived with \textsc{Period04}.}
\label{periodanalysis}
\end{center}
\end{figure}

\begin{figure}
\begin{center}
\includegraphics[width=0.47\textwidth]{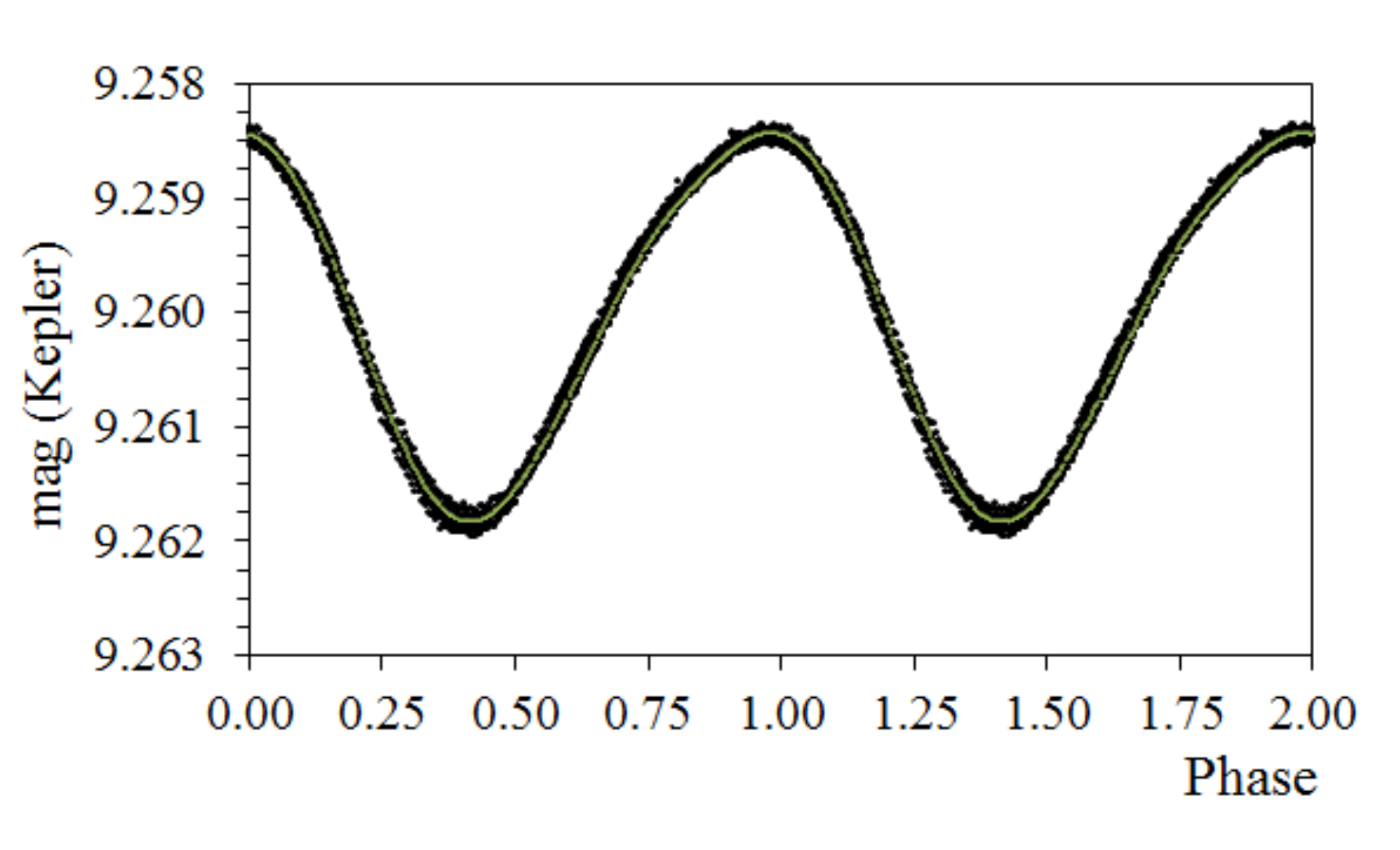}
\caption{Phase plot of KIC\,6128830, based on part of Q6 data ($575<(BJD-2454833)<590$) and folded with the main frequency of $f_1$\,=\,0.2065424\,d$^{-1}$. The fit curve is shown in green.}
\label{phaseplot}
\end{center}
\end{figure}

\kepler\ data for KIC\,6128830 are indicative of a single significant and independent frequency, $f_1$\,=\,0.2065424\,d$^{-1}$, boasting a S/N ratio of $\sim$186 and a peak-to-peak amplitude of $\sim$3.4\,mmag. In addition to that, harmonic frequencies up to $5f_1$ are present. Our results, which are in complete agreement with the findings of \citet{balona11} (cf. in particular their Fig. 10), are presented in Table \ref{pa_table} and Fig. \ref{periodanalysis}. A phase plot illustrating the light variations with the fundamental frequency is provided in Fig. \ref{phaseplot}. The fit curve, which has been calculated using the frequencies, amplitudes and corresponding phases derived with \textsc{Period04} (cf. Table \ref{pa_table}), adequately describes the observed light changes.

Several other periodogram features deserve note. First of all, there is a forest of frequencies with low amplitudes ($\sim$0.06\,mmag) around 0.175\,d$^{-1}$. The four most significant frequencies are, in order of decreasing significance, 0.173894\,d$^{-1}$, 0.176204\,d$^{-1}$, 0.174929\,d$^{-1}$ and 0.171929\,d$^{-1}$. Although we are unsure about their origin, we regard them as spurious detections that have likely been caused by instrumental artefacts or some other related phenomenon that has injected noise into the dataset. As indeed cautioned by several authors, the low-frequency features of \kepler\ data are poorly understood and frequencies below $\sim$0.2\,d$^{-1}$ should be regarded with caution \citep[e.g.][]{balona11}.

In addition to that, residual peaks of the fundamental frequency and its second harmonic are observed in the frequency spectra, which means that even after prewhitening for these frequencies, residual power is observed in the direct vicinity of the main peaks. As an example, the green arrow in panel (c) of Fig. \ref{periodanalysis} indicates the position of the residual peaks of the main frequency $f_1$. A detailed view of the corresponding part of the amplitude spectrum is provided in the inset.

\begin{figure}
\includegraphics[width=0.5\textwidth]{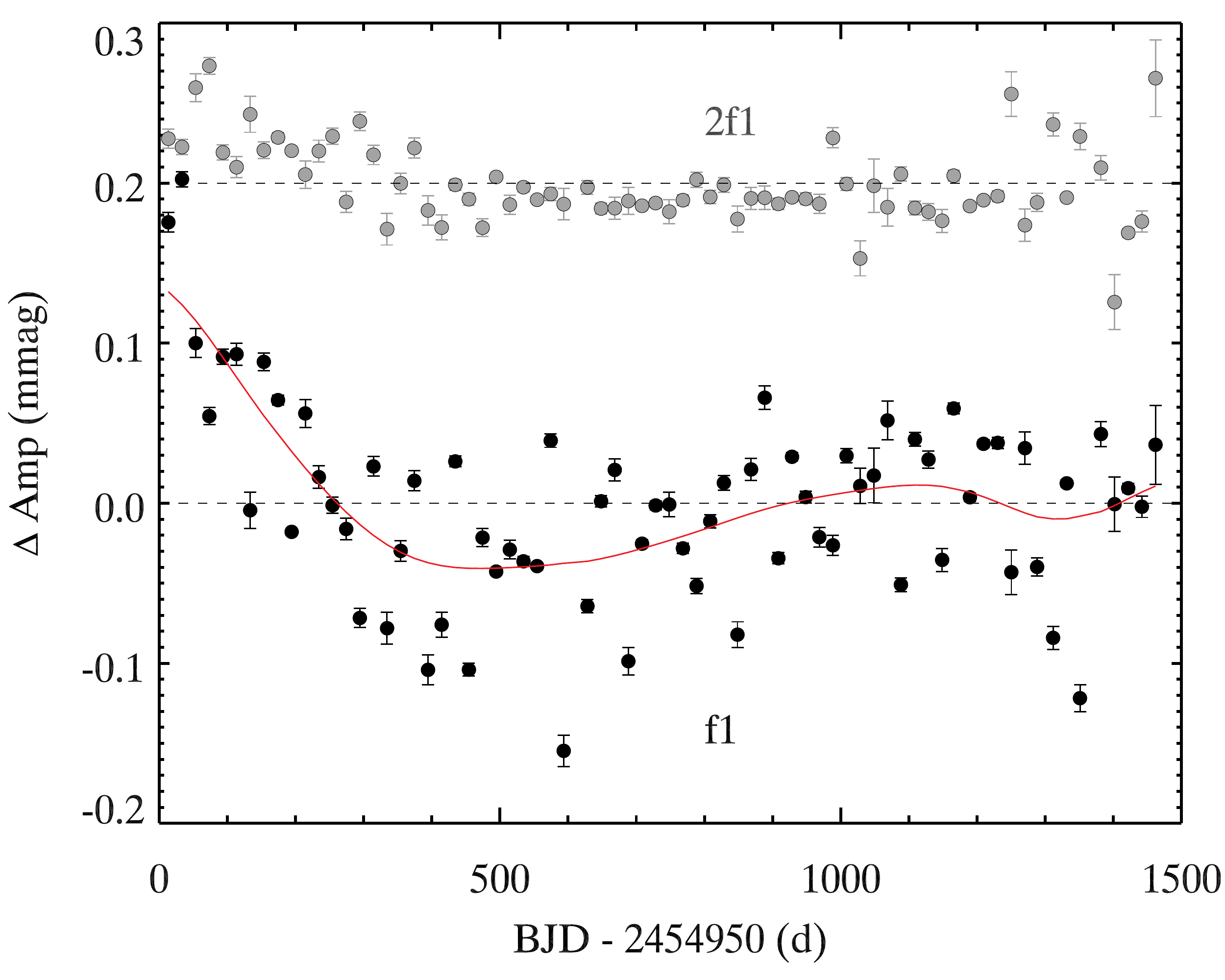}
\caption{Amplitude versus time of the fundamental frequency $f_1$ (black dots) and its second harmonic $2f_1$ (grey dots). Mean values have been subtracted, and an offset of 0.2\,mmag was employed for clarity. The solid red line is a Gaussian smoothing function applied to the $f_1$ data to highlight longer time-scale variation.}
\label{amp_mod}
\end{figure}

We attribute the observed residual peaks to amplitude modulation of the photometric variability, which is clearly seen in the light curve and also visible in Fig. \ref{lightcurve}. In order to investigate this phenomenon in more detail, a numerical analysis was performed. The data were divided into 20-day segments and the amplitude of the fundamental frequency $f_1$ and its second harmonic $2f_1$ were measured in each segment. The results are shown in Fig. \ref{amp_mod}. The amplitudes of $f_1$ and $2f_1$ change by about 10\% over the four years of \kepler\ coverage, but there is also additional large scatter on shorter timescales. Because a component of the amplitude modulation occurs on timescales longer than the time base of the data, and the observed shorter-timescale modulation is aperiodic, the amplitude modulation is not resolved into a triplet in the frequency spectra but results in the observed \lq messy\rq \,residuals instead.

\section{Evolutionary status} \label{evolutionarystatus}

\begin{figure}
	\includegraphics[width=0.47\textwidth]{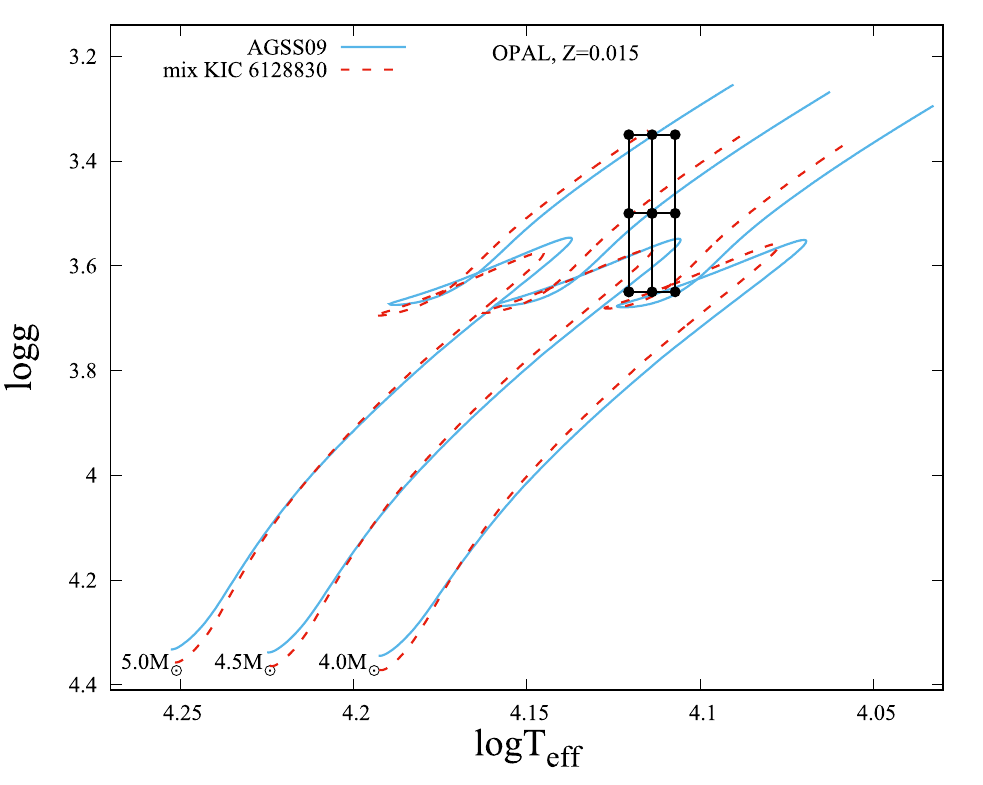}
  \caption{Kiel diagram indicating the location of KIC\,6128830 and the corresponding error box. Models were calculated for metallicity $Z=0.015$, H abundance $X=0.7$ and He abundance $Y=0.285$ and assuming solar composition (\citetalias{AGSS09}; solid blue lines) and the observed peculiar composition (mixKIC6128830; dashed red lines). See text for details.}
  \label{fig_evolution1}
\end{figure}

Evolutionary models were calculated with the MESA code \citep{MESA,MESA2,MESA3}, and the MIST configuration files \citep{MIST0,MIST1} were used in the computations. For all models, the OPAL opacity tables \citep{OPAL} have been employed. We included exponential overshooting from the hydrogen-burning convective core with the parameter $f_{\rm{ov}}=0.016$ \citep{Herwig2000}. Rotational velocities at Zero Age Main Sequence (ZAMS) were of the order of 36-39\,\kms, depending on the model.

Fig.\,\ref{fig_evolution1} illustrates the \logg\ vs. \teff\ (Kiel) diagram indicating the position of KIC\,6128830 and the corresponding error box. Metallicity $Z=0.015$, initial H abundance $X=0.7$ and He abundance $Y=0.285$ were assumed. Models have been calculated for masses $M=4.0\,M_{\odot}$, $M=4.5\,M_{\odot}$ and $M=5.0\,M_{\odot}$. Rotational velocities at the surface were adjusted to be consistent with the observed value of \vsini\,=\,25\,$\pm2$\,\kms. Models were calculated employing the \citetalias{AGSS09} mixture of chemical elements (solid blue lines) and the peculiar composition of KIC\,6128830 (mixKIC6128830; dashed red lines), as determined via our abundance analysis (cf. Section \ref{abundances}). In the case of missing information for some elements (e.g. N, Ar, Cl), we resorted to \citetalias{AGSS09} values in the calculation of the mixKIC6128830 models. As can be clearly seen, the effect of chemical composition is quite significant.

\begin{figure}
	\includegraphics[width=0.5\textwidth]{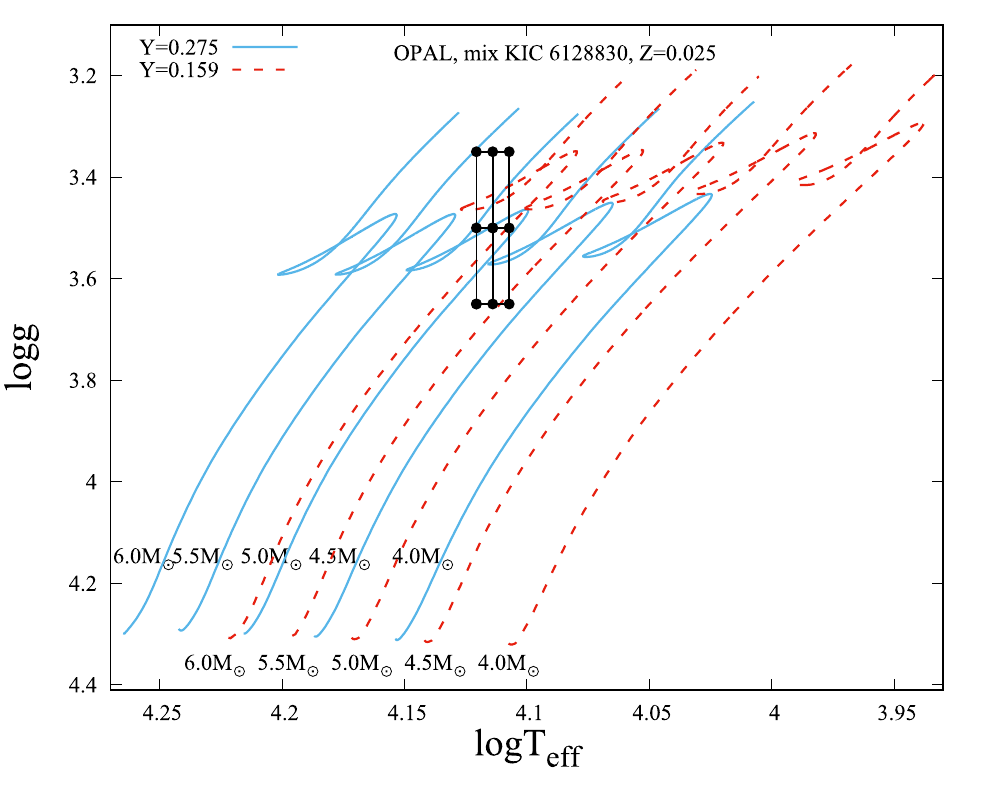}
  \caption{The same as in Fig.\,\ref{fig_evolution1}, but for metallicity $Z=0.025$ and He abundance $Y=0.275$ (solid blue lines) and $Y=0.159$ (dashed red lines).}
  \label{fig_evolution3}
\end{figure}

From our spectroscopic abundance analysis, we have derived $X=0.816$, $Y=0.159$ and $Z=0.025$. Evolutionary tracks calculated with this chemical composition are indicated by the dashed red lines in Fig.\,\ref{fig_evolution3}. As expected for a CP3 star, the derived He abundance is low, and we have explored models based on different values of $Y$ in order to study the He abundance effect. In Fig.\,\ref{fig_evolution3}, we also plotted models calculated with $Y=0.275$, spectroscopically-determined metallicity $Z=0.025$, and standard hydrogen abundance $X=0.7$ (solid blue lines). It can be seen that evolutionary tracks for lower values of $Y$ are significantly shifted towards lower effective temperatures, which in turn yields much higher masses for KIC\,6128830. In addition, a lower He abundance prolongs the main-sequence evolutionary phase. For an assumed standard chemical compositions ($X=0.7$, $Y=0.285$, $Z=0.015$, Fig.\,\ref{fig_evolution1}), we derive a mass of about 4.5\,$M_{\odot}$, while the spectroscopically-determined values ($X=0.816$, $Y=0.159$ and $Z=0.025$) indicate a mass of the order of 6\,$M_{\odot}$. Other mixtures give intermediate masses.

The differences in He abundance from spectroscopy and our models can be explained by atomic diffusion. The spectroscopically derived chemical abundances correspond to photospheric values that have been significantly modified; He has gravitationally settled towards the core, leaving an observed underabundance at the surface. The internal composition of the star, however, is different and should not deviate much from primordial composition, i.e. the He abundance will be higher and in keeping with big-bang nucleosynthesis and galactic chemical evolution. Obviously then, considerable uncertainty surrounds $Y$ and the derived mass but we strongly favour a value of 4.5\,$M_{\odot}$. In order to encompass all possible situations, we have considered different chemical compositions during the stellar pulsation analysis (cf. Section\,\ref{pulsationalmodels}).

We conclude that the star is most likely near the end of the main-sequence evolutionary phase. The mass deduced from the Kiel diagrams depends significantly on the assumed He abundance and varies from about $4.5M_{\odot}$ for $Y=0.275$ to about $6M_{\odot}$ for $Y=0.159$. We strongly favour the former mass value.

\section{Pulsation models} \label{pulsationalmodels}

In order to check whether the observed variability frequency $f_1$ can be explained by g-mode pulsation typical of an SPB star, we calculated pulsation models. We used the recently developed non-adiabatic pulsation code written by one of us (PW), which as yet remains unpublished and uses the traditional approximation of rotation \citep{1997ApJ...491..839L,2003MNRAS.340.1020T,2003MNRAS.343..125T,2007MNRAS.374..248D,2007AcA....57...11D}. Differential rotation was also included in our computations. The rotational velocity profiles were provided by the MESA evolution code, which takes into account angular momentum transport (treated in a diffusion approximation, \citealt{1978ApJ...220..279E,1989ApJ...338..424P,2000ApJ...528..368H}).

Some differences with respect to the models of \citet{turcotte03} and \citet{alecian09} deserve note. Whereas the models employed by these investigators include the effects of atomic diffusion and radiative levitation, our models consider the effects of diffusion but not radiative levitation. Including radiative levitation may slightly increase the instability of the calculated modes through the accumulation of iron (and iron group elements) in the pulsational driving zone. Furthermore, \citet{alecian09} used the \citet{dziembowski77} code, which is only partially nonadiabatic. While the outer layers are calculated fully nonadiabatically, the central regions are modeled through quasiadiabatic approximation (which is, of course, appropriate in the case of main-sequence stars). In contrast, our calculations are fully nonadiabatic in the entire star model. As pointed out above, we also use the traditional approximation of rotation and consider differential rotation. The resulting effect on the models will, however, be small since our object of interest rotates slowly. In summary, except for the handling of radiative levitation, we do not see any significant differences and expect that our results are qualitatively comparable to the computations of the aformentioned investigators.

\begin{figure}
  \includegraphics[width=0.5\textwidth]{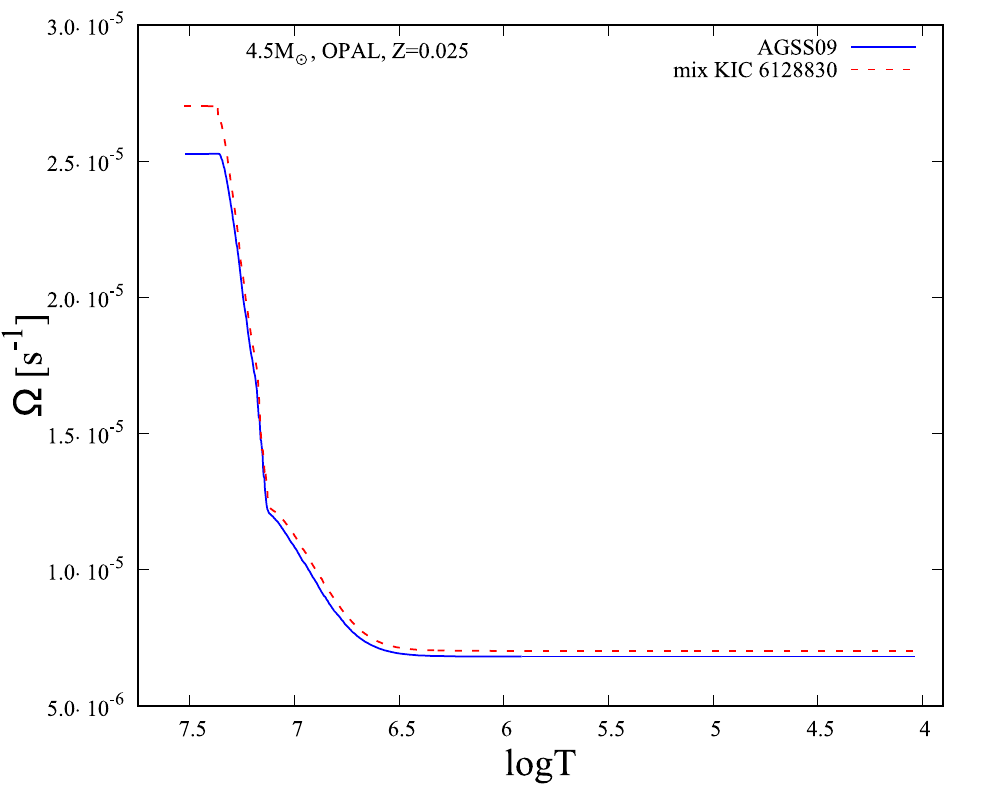}
  \caption{Rotational velocity profile of the evolutionary models calculated with $Z=0.025$, $M=4.5M_{\odot}$ and two mixtures of chemical composition: \citetalias{AGSS09} (solid blue line) and mixKIC6128830 (dashed red line).}
  \label{fig_pulsation1}
\end{figure}

The typical rotational velocity profile of our models is shown in Fig.\,\ref{fig_pulsation1}, where we plot the rotational frequency (in units s$^{-1}$) as a function of temperature for models calculated with the two different chemical compositions employed above (\citetalias{AGSS09} and mixKIC6128830). Our predictions indicate that the core of the star rotates much faster than the envelope, which has an important effect on pulsation models.

\begin{figure*}
  \includegraphics[width=0.5\textwidth]{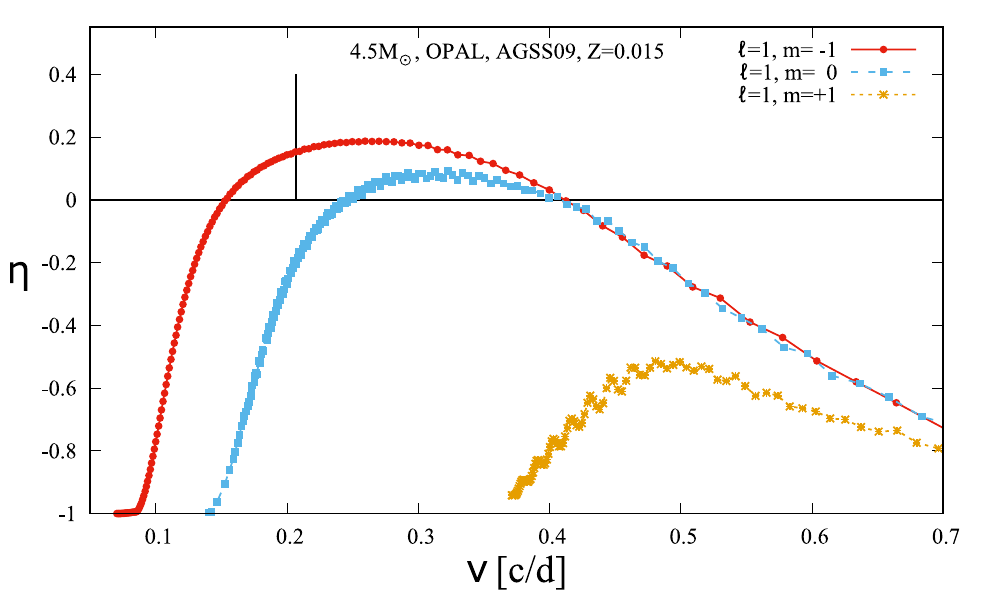}
  \includegraphics[width=0.5\textwidth]{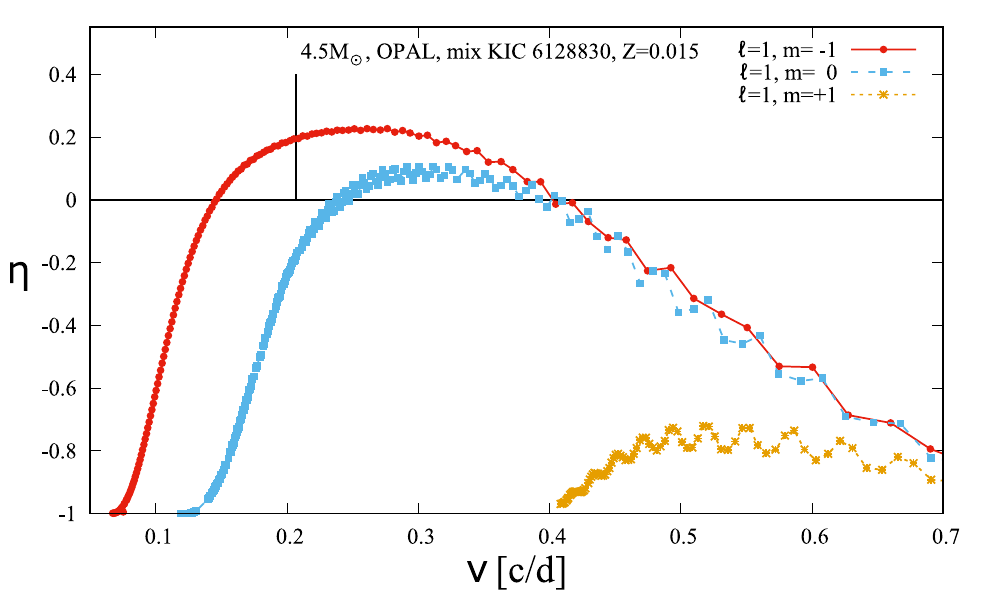}
  \caption{The instability parameter $\eta$ as a function of the pulsational frequency $\nu$. The vertical line indicates the observed variability frequency. The pulsational stability properties for the investigated dipole modes ($\ell=1$) with various azimuthal numbers ($m=-1$, $m=0$ and $m=1$) are shown. Left and right panels illustrate models calculated with, respectively, solar chemical composition (\citetalias{AGSS09}) and the observed peculiar composition of KIC\,6128830 (mixKIC6128830).}
  \label{fig_pulsation2}
\end{figure*}

Fig.\,\ref{fig_pulsation2} illustrates the normalized instability parameter, $\eta$, as a function of the frequency $\nu$. The $\eta$ parameter measures the net energy gained by a mode during the pulsational cycle \citep{stellingwerf78}. If $\eta>0$, then the corresponding mode is excited in a model. We have investigated dipole modes ($\ell=1$) with various azimuthal numbers, $m=-1$, $m=0$ and $m=1$, which indicate, respectively, the retrograde mode, the axisymmetric mode and the prograde mode. The vertical line indicates $f_1$. A solar chemical composition (\citetalias{AGSS09}) has been assumed for the models presented in the left panel of Fig.\,\ref{fig_pulsation2}, while the right panel illustrates the results using the peculiar composition of our target star (mixKIC6128830). Both models have been calculated assuming metallicity $Z=0.015$, mass $M=4.5M_{\odot}$ and similar values of $\log{T_{\rm{eff}}}\sim4.114$ and $\log{g}\sim3.6$. We note that the adopted value of $\log{g}$ is 0.1\,dex higher than the spectroscopically-derived value. It was chosen because the star is most likely near the end of the main-sequence evolutionary phase, which ends at $\log{g}\sim3.6$ for these models. The radii of models are $R=5.56R_{\odot}$ (\citetalias{AGSS09} composition) and $R=5.75R_{\odot}$ (mixKIC6128830 composition). There is a small difference in evolutionary phase between the models. The later model finds the star situated at the Terminal Age Main Sequence (TAMS), while the former is just near that phase (see Fig.\,\ref{fig_evolution1}). As can be clearly seen, the highest instability occurs for retrograde modes, followed by a strong instability for the axisymmetric modes; prograde modes are effectively damped by the differential rotation. We can see obvious mode trapping caused by the peculiar chemical composition and angular momentum gradients near the convective core.

\begin{figure}
  \includegraphics[width=0.5\textwidth]{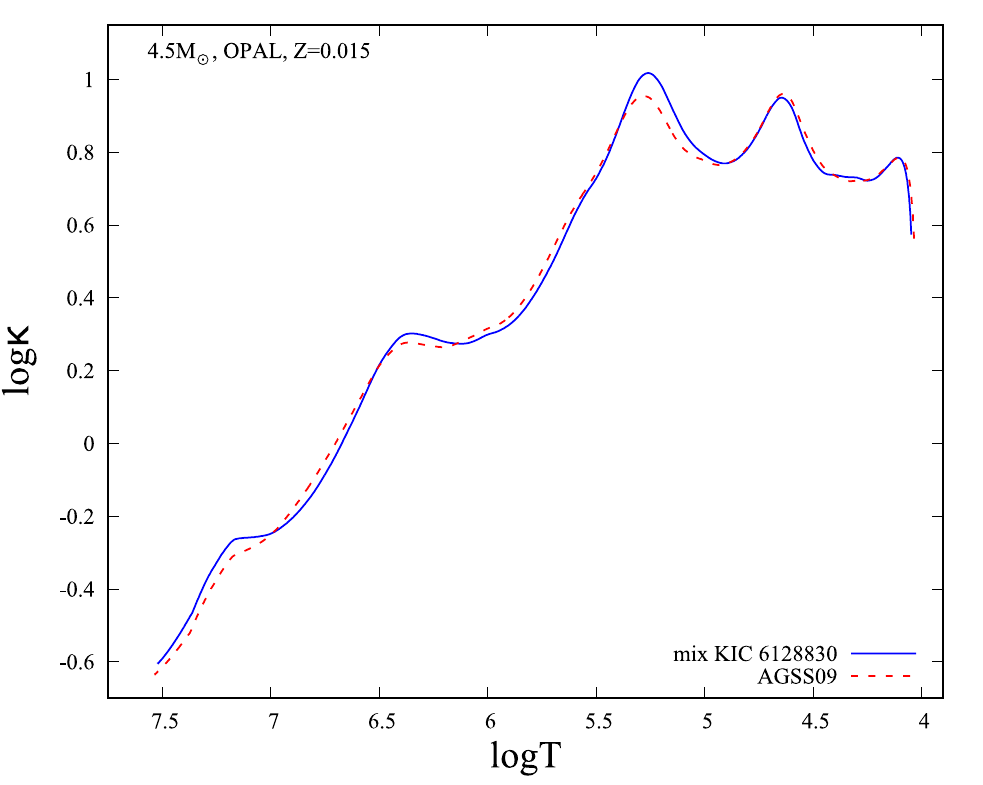}
  \caption{Comparison of the opacity coefficients for the models presented in Fig.\,\ref{fig_pulsation2}.}
  \label{fig_pulsation3}
\end{figure}

The intrinsic peculiar mixture of chemical elements observed in KIC\,6128830 slightly increases the instability parameter of retrograde and axisymmetric modes. This is caused mainly by the relatively high abundances of Fe and Mn, which effectively raise the opacity in the driving zones. In Fig.\,\ref{fig_pulsation3}, a comparison of the opacity coefficients for the models presented in Fig.\,\ref{fig_pulsation2} is given. As has been pointed out above, the prograde modes are damped more strongly in the model calculated with the mixKIC6128830 abundances because of the higher rotational velocity of the core (see Fig.\,\ref{fig_pulsation1}).

\begin{figure*}
  \includegraphics[width=0.5\textwidth]{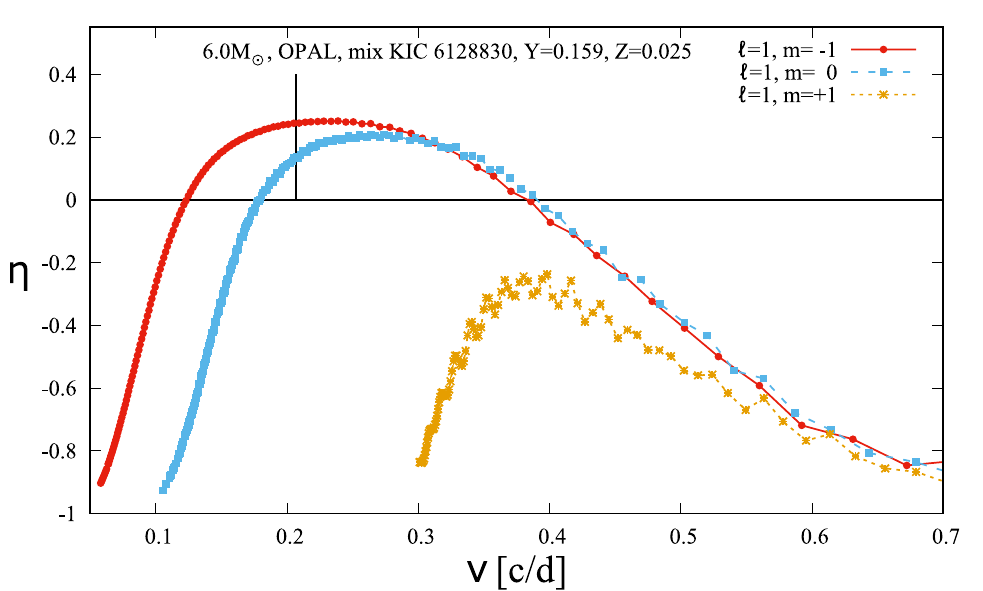}
  \includegraphics[width=0.5\textwidth]{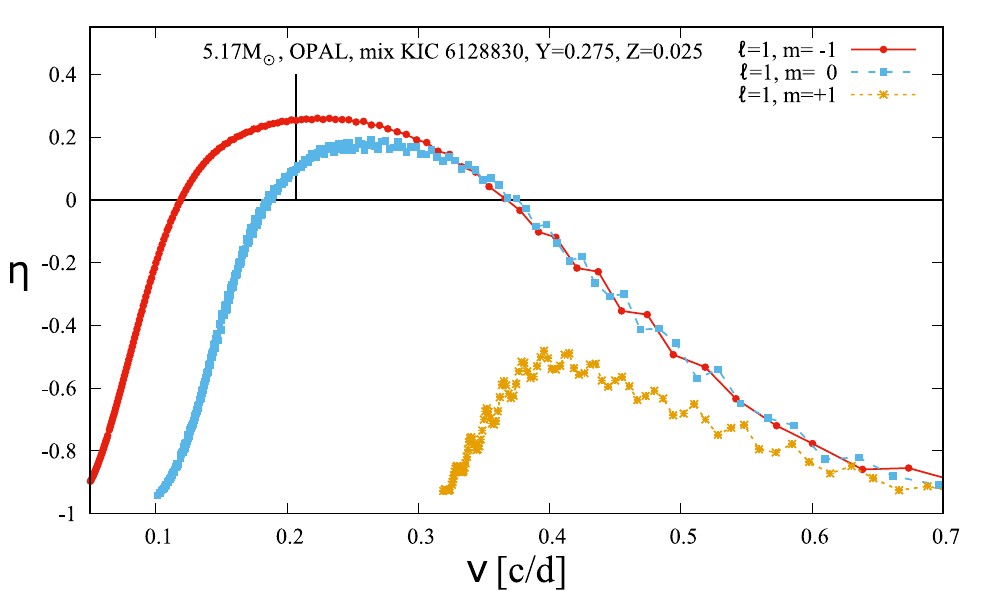}
  \caption{The same as in the right panel of Fig.\,\ref{fig_pulsation2}, but for $Z=0.025$, $Y=0.159$ (left panel) and $Y=0.275$ (right panel).}
  \label{fig_pulsation4}
\end{figure*}

The observed, low abundance of He also has significant effect on pulsation models. In the left panel of Fig.\,\ref{fig_pulsation4}, we have plotted the instability parameter for the model calculated with mixKIC6128830, He abundance $Y=0.159$, H abundance $X=0.816$ and metallicity $Z=0.025$. Model mass and radius are $M=6.0M_{\odot}$ and $R=6.99R_{\odot}$, respectively. It becomes obvious that the range of unstable modes is wider than in case of the models shown in Fig.\,\ref{fig_pulsation2}. Furthermore, the maximum value of the instability parameter is higher. In order to study the He abundance effect, we have investigated a model with the same metallicity, i.e. $Z=0.025$, but with higher He abundance $Y=0.275$ and standard hydrogen abundance $X=0.7$. Model mass and radius are $M=5.17M_{\odot}$ and $R=6.70R_{\odot}$, respectively. This model is shown in the right panel of Fig.\,\ref{fig_pulsation4}. Both models occupy similar positions in the Kiel diagram with $\log{T_{\rm{eff}}}\sim4.114$ and $\log{g}\sim3.6$. The main difference lies in the values of the $\eta$ parameter of the prograde modes, which are lower in the case of higher He abundance. Again, this is caused by the faster rotation of the convective core in the model with increased He abundance.

Since there is just one independent frequency, which -- anticipating our results -- we interpret as being due to a non-pulsational origin, we decided to search for a metallicity value at which pulsation would be suppressed. It turns out that this applies only for a metallicity smaller than $Z=0.005$. This is a rather small value, more appropriate for the Magellanic Clouds than for a star from the Milky Way.

In summary, our models predict that g-mode pulsational instability occurs both in the case of assuming standard solar chemical composition (\citetalias{AGSS09}) as well as the peculiar composition of KIC\,6128830 (mixKIC6128830) and exists for a wide range of metallicities.

\section{Discussion} \label{conclusion}

The present work unambiguously identifies KIC\,6128830 as a bona-fide CP3 star (cf. Section \ref{abundances}). As has been pointed out, very few examples of photometric variations in CP3 stars are known from ground-based observations. This picture is now slowly changing, and the advent of space-based photometry provides the opportunity for the first precise characterizations of variability in this group of stars. However, there is no concensus yet as to what causes the observed light changes in CP3 stars. Pulsational variability has been postulated from theoretical considerations, but recent studies have favoured rotational modulation as the underlying mechanism of the observed variations (cf. Sect. \ref{introduction}). However, no concensus has as yet been reached.

In the case of our target star, arguments can be made in favour of either interpretation. Temperature-wise, KIC\,6128830 falls clearly into the SPB instability domain \citep{2017MNRAS.469...13S}. The star is in a rather advanced evolutionary state, though, and the errors on $T_{\rm{eff}}$ and $\log{g}$ are non-negligible (cf. e.g. Fig. \ref{fig_evolution1}). Calculations indicate that the SPB instability strip generally ends close to the TAMS \citep{2017MNRAS.469...13S}. However, assuming our target is a main-sequence star, it should still lie well within the SPB instability domain. More importantly, our calculations indicate the occurrence of g-mode pulsations for a wide range of input parameters (Sect. \ref{pulsationalmodels}). The observed variability frequency $f_1$\,=\,0.2065424\,d$^{-1}$ is situated in the middle of the theoretical frequency instability range. From theoretical considerations, we therefore expect g-mode pulsation typical of an SPB star in KIC\,6128830.

To distinguish rotational and pulsational variability, it is helpful to investigate the frequency spectra of the corresponding stars. Many pulsating variables, like the SPB or $\gamma$ Dor stars, exhibit multiple independent periods and quite different frequency spectra from rotating variables. For instance, harmonics of pulsation modes are expected only when the amplitude is large, and large-amplitude variability tends to coincide with multiperiodicity. On the other hand, harmonics are a consequence of localized spots and therefore a characteristic of the frequency spectra of rotating variables (\citealt{balona15}; cf. also \citealt{huemmerich16}, in particular their Fig. 3).

Only one significant and independent frequency has been identified in our target star, which exhibits strictly monoperiodic variability throughout the four years of \kepler\ coverage. In addition, corresponding harmonics up to $5f_1$ are present (cf. Section \ref{section_periodanalysis}). It is hard to reconcile this result with the presence of pulsational variability; in fact, the observed frequency spectrum of KIC\,6128830 (Fig. \ref{periodanalysis}) is decidedly not like that of a pulsational variable.

SPB stars, in particular, are notoriously multiperiodic. In their investigation of ten new SPB star candidates discovered in \textit{Hipparcos} data, \citet{mathias01} identify monoperiodic light variability in one star of their sample, HD\,28114. However, this result has been based on 56 measurements only, and the authors caution that, quite probably, additional frequencies will be revealed in photometry of higher precision. Other monoperiodic SPB candidates were later reclassified as CP stars, i.e. rotational variables \citep{decat01}. The advent of ultra-precise photometry from space (as provided by e.g. the \textit{CoRoT} and \kepler\ missions) has revolutionized the field, and multiple frequencies are routinely found for SPB variables \citep{balona11,mcnamara12}. In fact, the presence of multiple frequencies is commonly used as a criterion to distinguish SPB stars from rotating variables \citep[e.g.][]{mcnamara12}. In the aforementioned studies, all SPB stars identified in \kepler\ data exhibit multiple frequencies, often also in the $\beta$ Cephei star realm, in which case they are classified as SPB/$\beta$ Cep hybrid pulsators. To sum up, we are unaware of the existence of any monoperiodic SPB variables. The observed \vsini\ value and the model radii are compatible with the rotational modulation scenario. Furthermore, the strictly monoperiodic nature of the photometric variability in KIC\,6128830, as well as the presence of multiple harmonics in the frequency spectrum, provide strong evidence in favour of a rotational origin. This is in line with the results of several recent investigations that have identified monoperiodic variability in CP3 stars, too (cf. Section \ref{introduction}). 

Interestingly, amplitude modulation on short and long timescales is present in the light curve of KIC\,6128830, the interpretation and origin of which remains unclear. In Section \ref{section_periodanalysis} we showed that the amplitudes of $f_1$ and $2f_1$ change by about 10\% over the four years of \kepler\ coverage. Additional, aperiodic scatter on shorter timescales is also present.

The short-term amplitude modulation is clearly seen in the light curve and certainly not the result of an error in our analysis. However, spot sizes are definitely not expected to modulate on timescales of days, and we have not yet identified the physical cause of this puzzling phenomenon.

The long-term amplitude modulation is obvious in the first $\sim$300 days of \kepler\ coverage, after which it becomes less pronounced. In order to investigate its reality, we (re)calculated our results using both our detrended data and data processed with the msMAP pipeline and obtained at the Kepler Asteroseismic Science Operations Centre (KASOC)\footnote{http://kasoc.phys.au.dk/}. The trend is seen in both data sources. We can therefore rule out that our detrending procedure is at the root of the observed long-term changes.

We cannot exclude that it is due to systematic effects in the \kepler\ data, although -- to the best of our knowledge and according to our experience -- there is no artefact causing amplitude modulation on these timescales. Changes in the amount of flux captured in the aperture might be a viable explanation. This, however, usually happens on the very different timescale of one \kepler\ quarter ($\sim$91\,d), and is corrected for quite well in the msMAP reductions. Furthermore, the pulsation amplitude is not a monotonically decreasing function of time (Fig. \ref{amp_mod}), nor is it periodic with the Kepler roll period, which together argue against CCD deterioration as the cause of any amplitude modulation.

Building on the available evidence, we therefore favour the hypothesis that the long-term amplitude modulation is intrinsic to the variability pattern of the star. Several studies (e.g. \citealt{kochukhov07}) have reported secular evolution of the abundance inhomogeneities observed in CP3 stars, and it is intriguing to surmise that, perhaps, the observed long-term amplitude changes are in fact the photometric manifestation of this phenomenon. At \kepler\ precision, this certainly seems plausible. Of course, only long-term photometric and spectroscopic monitoring will be able to shed more light on the origin of the observed long-term amplitude changes.

\section{Conclusion}

We have carried out a detailed spectroscopic and photometric investigation of the candidate CP3 star KIC\,6128830. Our detailed abundance analysis confirms the star's status as a bona-fide CP3 star. We have investigated the photometric variability of our target star using the available 4-yr \kepler\ data. In agreement with the findings of \citet{balona11}, we have identified a single significant and independent frequency $f_1$\,=\,0.2065424\,d$^{-1}$ and harmonic frequencies up to $5f_1$.

We have discussed our results in detail, with the aim of investigating the origin of the observed light variations. Unfortunately, with the available material, we are unable to settle this question conclusively. In our opinion, the argument with the most weight is the completely monoperiodic nature of the light variations, which is typical of a rotational variable and not to be expected in a pulsating star - in particular not in an SPB star, which have proven to be multiperiodic objects in all other \kepler\ data. We therefore favour a rotational origin of the observed light changes. Phase-resolved high-resolution spectroscopy is needed to settle this question conclusively, which we herewith strongly encourage. The resolution of this issue is of considerable interest and will provide constraints and tests for pulsation models as well as contribute to the understanding of the nature of CP3 stars and their photometric variability.

\section*{Acknowledgements}
We thank the referee for his/her thoughtful report which helped to significantly improve the paper. EN acknowledges the Polish National Science Centre grants no. 2014/13/B/ST9/00902. PW was supported by the Polish National Science Centre grants no. DEC-2013/08/S/ST9/00583. This work was also supported by the grant 7AMB17AT030 (M\v{S}MT). Calculations have been carried out at the Wroc{\l}aw Centre for Networking and Supercomputing (http://www.wcss.pl), grants No.\,214 and No.\,265. This paper includes data collected by the \kepler\ mission. Funding for the \kepler\ mission is provided by the NASA Science Mission directorate. Some of the data presented in this paper were obtained from the Mikulski Archive for Space Telescopes (MAST). STScI is operated by the Association of Universities for Research in Astronomy, Inc., under NASA contract NAS5-26555. Support for MAST for non-HST data is provided by the NASA Office of Space Science via grant NNX09AF08G and by other grants and contracts. We are grateful to the \kepler\ team for giving public access to \kepler\ light curves.


\bibliographystyle{mnras}
\bibliography{KIC6128830}

\bsp	
\label{lastpage}
\end{document}